\documentclass[a4paper, amsfonts, amssymb, amsmath, reprint, showkeys, nofootinbib, twoside, floatfix]{revtex4-1}
\pdfoutput=1
\usepackage[english]{babel}
\usepackage[utf8]{inputenc}
\usepackage[colorinlistoftodos, color=green!40, prependcaption]{todonotes}
\usepackage{amsthm}
\usepackage{mathtools}
\usepackage{physics}
\usepackage{xcolor}
\usepackage{graphicx}
\usepackage[left=23mm,right=13mm,top=35mm,columnsep=15pt]{geometry} 
\usepackage{adjustbox}
\usepackage{placeins}
\usepackage[T1]{fontenc}
\usepackage{lipsum}
\usepackage{csquotes}
\usepackage[pdftex, pdftitle={Article}, pdfauthor={Author}]{hyperref} 

\begin{document}


\title{Selected Machine Learning of HOMO-LUMO gaps with Improved Data-Efficiency }

\author{Bernard Mazouin}
    \affiliation{University of Vienna, Faculty of Physics, Kolingasse 14-16, 1090 Vienna, Austria}
\author{Alexandre Alain Sch\"opfer}
    \affiliation{Department of Chemistry, University of Basel, Klingelbergstrasse 70, 4056 Basel, Switzerland}
\author{O. Anatole von Lilienfeld}
    \email[Correspondence email address: ]{anatole.vonlilienfeld@univie.ac.at}
    \affiliation{University of Vienna, Faculty of Physics, Kolingasse 14-16, 1090 Vienna, Austria}
        \affiliation{Department of Chemistry, University of Basel, Klingelbergstrasse 70, 4056 Basel, Switzerland}

\date{\today} 

\begin{abstract}

Quantum Machine Learning (QML) models of molecular HOMO-LUMO-gaps often struggle to achieve satisfying data-efficiency as measured by decreasing prediction errors for increasing training set sizes. 
Partitioning training sets of organic molecules (QM7 and QM9-data-sets) into three classes [systems containing either aromatic rings and carbonyl groups, or single unsaturated bonds, or saturated bonds] prior to training results
in independently trained QML models with improved learning rates. 
The selected QML models of band-gaps (at GW, B3LYP, and ZINDO level of theory)  reach mean absolute prediction errors of $\sim$0.1 eV  for up to an order of magnitude fewer training molecules than for conventionally trained models.
Direct comparison to $\Delta$-QML models of band-gaps suggest that selected QML possesses substantially more data-efficiency.
The findings suggest that selected QML, e.g.~based on simple classifications prior to training, could help to successfully tackle challenging quantum property screening tasks of large libraries with high fidelity and low computational burden. 
\end{abstract}


\maketitle

\graphicspath{{ms_figures/}}

\section{Introduction} \label{sec:intro}

Machine Learning (ML) based surrogate models of quantum properties have gained a lot of traction in recent years \cite{Ramakrishnan2017, Lilienfeld2018, Lilienfeld2020, Lilienfeld2020a, Huang2021}. 
This rise in interest is partly driven by the computational efficiency of ML algorithms that typically outpace the conventional quantum chemistry methods which attempt to numerically solve sophisticated approximations to the electronic Schr\"odinger equation.
The application of these algorithms to Chemical Compound Space (CCS) is commonly referred to as Quantum Machine Learning (QML).
During training, QML models get parameterized in terms of a heuristic functional form which encodes a statistical relation between sample training molecules and their corresponding labels (quantum property). The resulting QML model can subsequently be used to make quantum property predictions throughout CCS, i.e.~for unknown out-of-sample molecules. 
Since its inception in 2012~\cite{Rupp2012}, QML has already been applied to a variety of chemical classes including, among others,  organic molecules \cite{ramakrishnan2015many, Faber2017, Stuke2019}, amino acids \cite{Stuke2019}, polymers \cite{Patra2020}, or solids \cite{faber2016machine, Pilania2017, Isayev2017, Li2018, Butler2018}. 
Within these applications, it has been used to predict ab initio thermodynamic properties such as atomization energies \cite{Rupp2012, Hansen2013, ramakrishnan2015many, Faber2017}, energy above convex hull \cite{Li2018}, or free energy of solvation \cite{Weinreich2021}, as well as electronic properties such as HOMO and LUMO energies or dipole moments \cite{Montavon2013, Ramakrishnan2015a, ramakrishnan2015many, Faber2017, Schuett2019, christensen2019operators}.
Some state-of-the-art QML models can  reach an accuracy on par with quantum chemistry algorithms already for modest training set sizes \cite{Faber2017}, and are thus well positioned for their direct application in computational materials design efforts \cite{Pyzer-Knapp2015, Shandiz2016, Gomez-Bombarelli2016, Sendek2019, Zunger2018, Jorgensen2018}.

\begin{figure}[h!]
    \includegraphics[width=\linewidth]{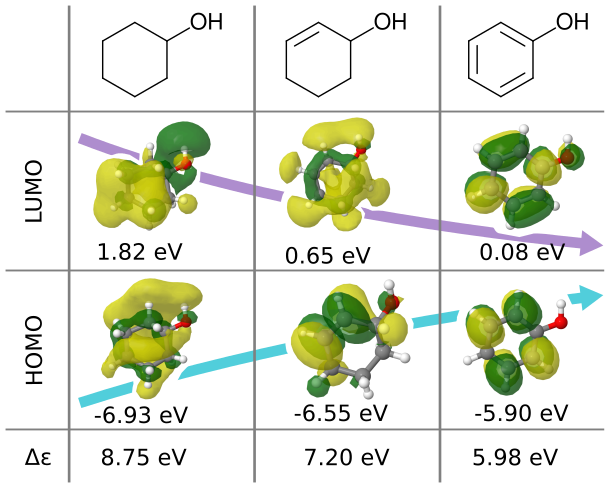}
    \caption{Illustration of frontier molecular orbitals and eigenvalues being dominated by simple features such as bond-saturation: Compositionally and structurally similar molecules (cyclohexanol, cyclohex-2-enol and phenol) exhibit vast differences in HOMOs, LUMOs, and eigenvalues. The orbitals are visualized with Jmol \protect{\cite{Jmol}} using results from B3LYP/6-31G(2df,p) calculations performed with ORCA 4.0.1 \protect{\cite{Neese2012}}.}
    \label{fig:drawing}
\end{figure}

Not surprisingly, the importance of rapid yet accurate QM property predictions has inspired the development of specialized ML methods.
For example, optimized representations or Neural Network architectures have been designed just for this purpose \cite{Behler2007, Montavon2013, Schuett2014, Faber2017, Pereira2017, Stuke2019, Schuett2019, Unke2019, Westermayr2020a}.
In particular, one can adjust the QML procedure to the property of interest by including more information about the underlying physics in it. 
In order to obtain QML models with higher data-efficiency for atomization energies, more descriptive representations such as SLATM \cite{Huang2020} and FCHL \cite{faber2018alchemical, Christensen2020}, which include 3-body-terms and physically motivated power laws, yield better results than the more heuristic CM \cite{Hansen2013} or BoB \cite{Hansen2015} representations, which merely encode the nuclear repulsion terms. 
The integration of gradients in KRR has led to reduced errors for response properties such as the dipole moment or forces \cite{Chmiela2017, Chmiela2018, christensen2019operators, Christensen2020a}. 
Furthermore, a biased selection of training samples will also lead to QML models with improved  accuracy \cite{Browning2017,Huang2020}.

\begin{figure}
    \includegraphics[width=\linewidth]{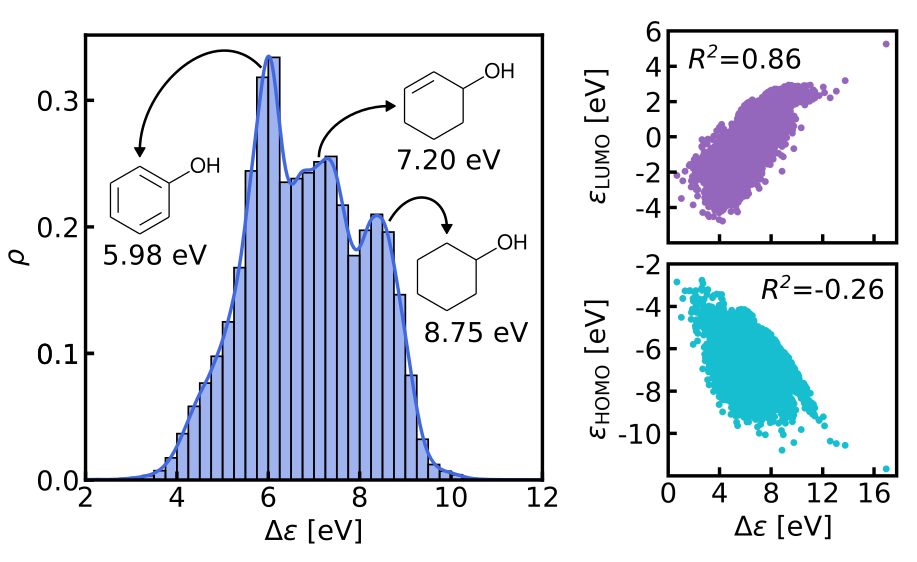}
    \caption{Left: Normalized histogram and kernel density estimate (solid line) of HOMO-LUMO gaps in QM9 data-set.
    Gaps of 3 similar molecules (cyclohexanol, cyclohex-2-enol and phenol) are indicated.
    Right: The HOMO and LUMO energies plotted against the HOMO-LUMO gaps with the respective correlation coefficients.}
    \label{fig:kde_qm9}
\end{figure}

Among the various QM properties frequently evaluated, the eigenvalues of the frontier orbitals, i.e.~highest occupied and lowest unoccupied molecular orbitals (HOMO and LUMO), are of special interest. 
These MO energies are intimately related to chemical reactions, polarizability, the optical gap and excitation energies. Their prediction often plays an important role for design decisions in the development of technological applications such as synthesis planning, electrochromic devices, light-emission diodes or photovoltaic solar panels \cite{Kubatkin2003, Roncali2007, Jurow2010, Beaujuge2010, Tao2017, Stoliaroff2020}. 
Interestingly, the generation of accurate QML models of frontier orbital eigenvalues proves more difficult than for other quantum properties---even when using molecular training sets of considerable size. 
Consequently, significant research efforts are currently being made in order to devise QML models of MO energies with improved data-efficiency.

We believe that this difficulty is partly, if not mostly, due to the intensive nature of MO energies.
Molecules with very similar stoichiometry and geometry do not necessarily have similar HOMO-LUMO gap values (see e.g.~the molecules  drawn in Fig.~\ref{fig:drawing}), whereas structurally dissimilar molecules can have very close values.
While the latter can be resolved easily by allowing for QML models which are not monotonic in CCS, 
the former point represents the actual challenge since all ML models are based on similarity arguments and smoothness assumptions.  
The lack of smoothness (as on display in Fig.~\ref{fig:drawing}) indicates the presence of additional dimensions, not properly being taken into account when relying on the conventional QML representations~\cite{Lilienfeld2020}. Inspection of HOMO-LUMO gap distributions (see Fig.~\ref{fig:kde_qm9}), however, indicates the superposition of multiple groups, possibly accounting for the `hidden' dimensions. 
In this work, we have investigated this hypothesis in more depth. 

\begin{figure}
    \includegraphics[width=\linewidth]{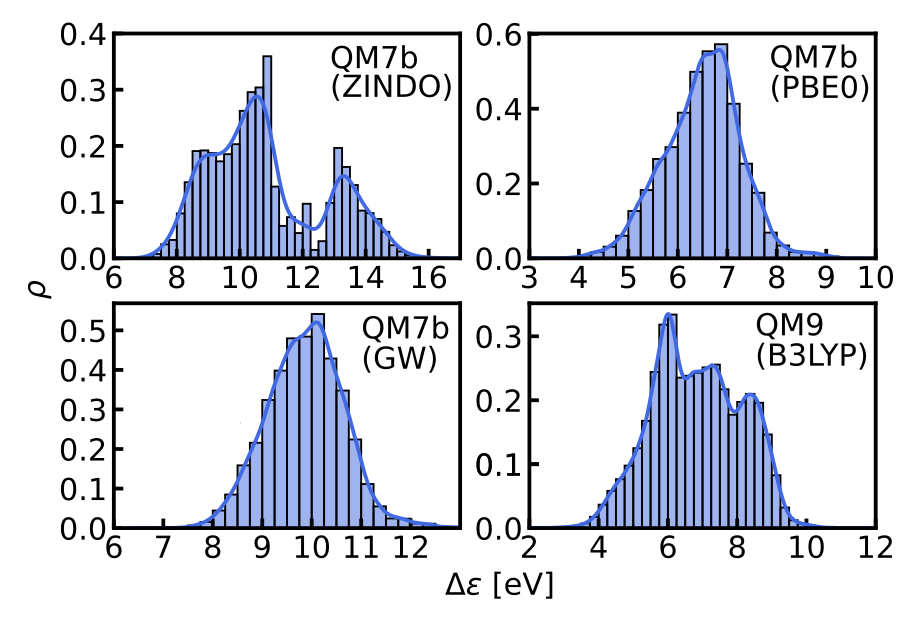}
    \caption{Histograms and KDEs of the HOMO-LUMO gaps of all molecules from QM7b at ZINDO, PBE0 and GW levels of theory and of all molecules from QM9 at B3LYP level of theory.}
    \label{fig:gaps_hist}
\end{figure}

We present the SML (Selected ML) method, a divide-and-conquer-like strategy which improves the data-efficiency of QML models of MO energies.
Prior to training, we partition the training data into smaller classes, and we train QML models separately for each class. 
The idea for such a classification is based on the peculiar shape of the distribution of HOMO-LUMO-gaps obtained from B3LYP in QM9 or ZINDO in QM7b: it is multi-modal  and appears to be composed of 3 sub-distributions, one per peak (see Fig.~\ref{fig:kde_qm9} and \ref{fig:gaps_hist}). 
According to frequency analysis the molecules can be easily classified into three groups, solely based on simple structural features.
The three example molecules indicated in Fig.~\ref{fig:kde_qm9}, that are each located close to a different peak of the distribution, indicate such features (saturated bonds, unsaturated bond, aromatic ring) that turn out to encode important information about the gap. 
Fig.~\ref{fig:drawing} also illustrates their dramatic effect on the character of their frontier orbitals, and thereby on eigenvalues and their gap.
Based on the QM9 analysis, we have defined a set of simple rules for classification which ({\em vide infra}) results in subsequently trained QML models, henceforth dubbed SML, with much improved learning curves.

 \section{Data and Methods} \label{sec:methods}

\subsection{Data}

The QM7b data-set \cite{Blum2009, Montavon2013} contains properties of 7211 organic molecules with up to 7 heavy atoms (C, O, N, S and Cl). 
These molecules were derived originally from the GDB-13 data-set. 
Thermodynamic and electronic properties are available at different levels of theory, which makes the data-set suitable for $\Delta$-ML applications \cite{Ramakrishnan2015}.
For this work, we use the HOMO and LUMO energies as obtained from ZINDO \cite{Ridley1973, Zerner1991} and GW \cite{Hedin1965, Aryasetiawan1998} calculation for direct and $\Delta$-ML. 
The HOMO-LUMO gaps correspond to the differences of LUMO and HOMO energies.

The QM9 data-set, published in 2014 by Ramakrishnan at al. \cite{Ruddigkeit2012, Ramakrishnan2014}, consists of more than 133k organic molecules with up to 9 heavy atoms (C, N, O and F) with corresponding geometries, thermodynamic and electronic properties.
These molecules were obtained from the GDB-17 data-set which contains over 166 billion molecular graphs. 
The properties were computed using DFT/B3LYP \cite{Hohenberg1964, Kohn1965, Becke1993a} with a 6-31G(2d,f) basis set. 
Over the last years, it has become an increasingly popular data-set in the QML community as it has been used as a staple to benchmark new QML models \cite{Gilmer2017, faber2018alchemical, Stuke2019, Schuett2018, Chen2019, Anderson2019, Lu2019, Liu2020}.

\subsection{Frequency Analysis and Classification}

\begin{figure*}
    \includegraphics[width=\textwidth]{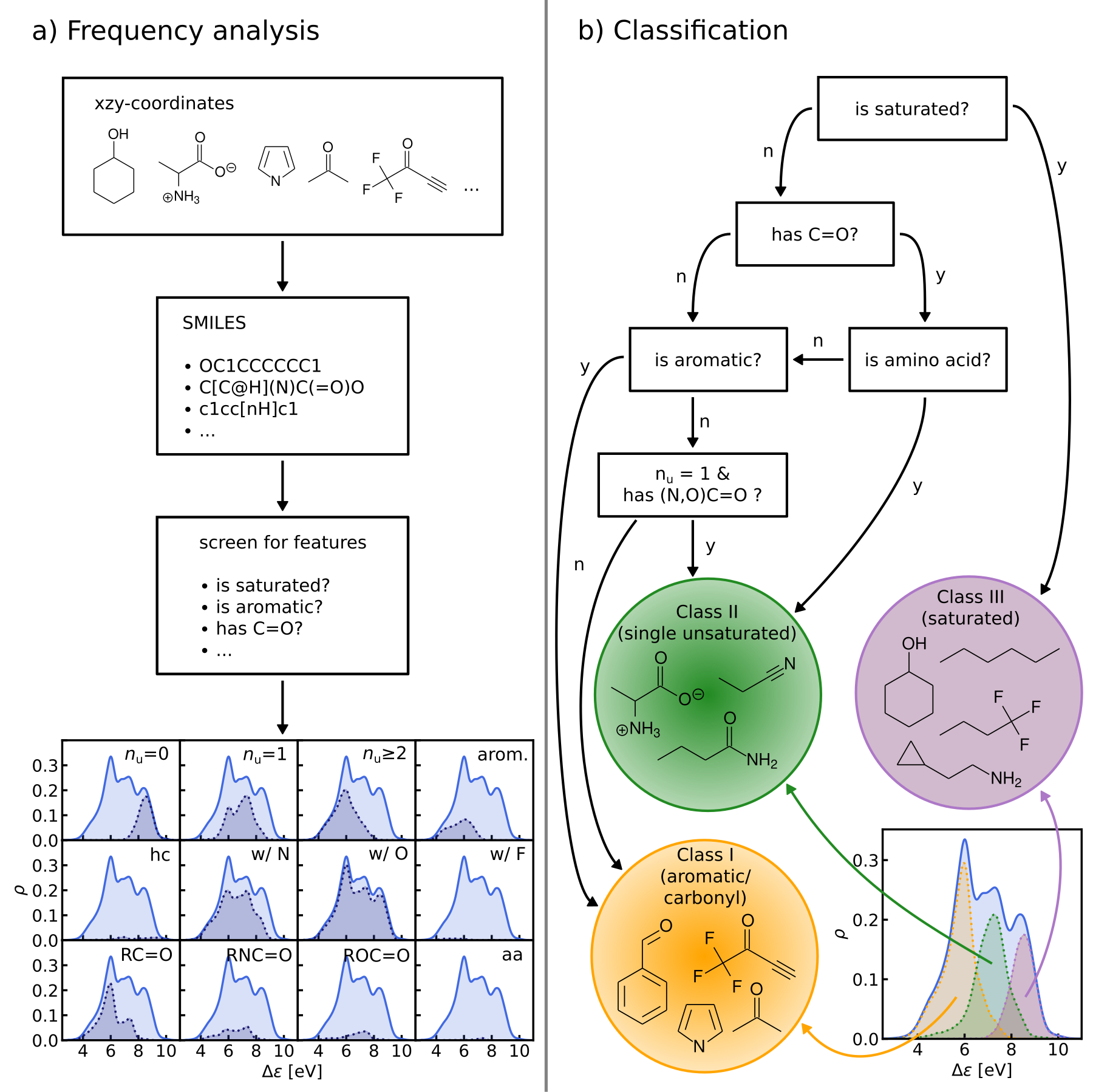}
    \caption{a) General procedure of the frequency analysis. 
    We use SMILES strings obtained from QM9 molecules to screen for features such as double bonds, aromatic rings, carbonyl groups and so on. 
    The plot at the bottom shows the results of this frequency analysis. 
    Each quadrant shows the distribution of HOMO-LUMO-gaps of a subgroup of QM9 matching a given criterion. 
    The top row shows the effect of saturation ($n_u$: number of unsaturated bonds, i.e. any double, triple or aromatic bond), the middle row shows the effect of elemental constitution (hc: hydrocarbons, w/ N: with nitrogen and so on) and the bottom row shows various kinds of carbonyl groups (columns 1 - 3) and amino acids (column 4). 
    b) Flowchart detailing the sequence of decisions that results in our final classification. 
    The distributions of the classes are highlighted in plot on the bottom left. }
    \label{fig:workflow}
\end{figure*}

We perform a frequency analysis to identify functional groups that relate to the HOMO-LUMO gap in our molecules (see Fig.~\ref{fig:workflow}, panel a)). 
By screening for a set of structural features and functional groups such as double bonds, aromatic rings or carbonyl groups e.g. using SMILES \cite{Weininger1988, Weininger1989, Weininger1990} strings and substructure matching as implemented in RDKit \cite{rdkit}, we tag the molecules in the data-set. 
For each tag, we compute the the Kernel Density Estimation (KDE) of the HOMO-LUMO-gaps of the matching molecules, normalize it with respect to the total number of molecules in the entire data-set, and draw it over its KDE. 
By visual inspection of the resulting plots we have detected those functional groups which govern the assignment to one of the classes in the  HOMO-LUMO-gap distribution.

In the next step, based on the frequency analysis, we define simple rules to separate the molecules into disjoint classes.
We make sure that the class distributions have a unimodal shape and coincide with the peaks of the total distribution (see Fig.~\ref{fig:workflow}, panel b)). 
For example, the distribution of saturated molecules in QM9 fits closely underneath the right peak, so that we can assign all saturated molecules to the class corresponding to that peak. 
The distribution of all carbonyl compounds however has two peaks that coincide with the left and middle peaks.
Therefore we have further subdivided the group of carbonyl compounds by distinguishing, for instance, between those with aromatic rings from those without, until one ends up with unimodal subdistributions.

\subsection{Kernel Ridge Regression}

The main idea behind supervised learning is to establish and exploit statistical relations between inputs $\mathbf{X}_i$ and corresponding target property label outputs $y_i$. 
In our case, the inputs are molecular representations which, in strict correspondence to Schr\"odinger's equation, encode stoichiometry and  geometry.
We have relied on the SLATM representation \cite{Huang2020}, which describes a molecule as a spectrum of atomic, 2-body and 3-body terms. 
The target labels are the properties of interest, i.e. the HOMO-LUMO gaps and the individual frontier orbital energies.
A training set $\{\mathbf{X}_i, y_i\}_{i=1}^{N_{\mathrm{tr}}}$ is a sample for which both the inputs and target values are known, whereas for the test set only the inputs $\{\mathbf{X}_j\}_{j=1}^{N_{\mathrm{te}}}$ are known, but the target values unknown. 
The ML model uses the training data to infer a statistical model that relates the input $\mathbf{X}_i$ to the output $y_i$. 
This statistical model can then be applied to the molecules in the test set in order to produce a prediction error estimate for the corresponding properties. 
As such, ML circumvents numerically solving the Schr\"odinger equation, and provides instead statistical estimates which are computationally more efficient than state-of-the-art quantum chemistry calculations.

We are dealing with a regresssion problem where the task is to predict continuous target values. 
Our method of choice is KRR \cite{Muller2001, Schoelkopf2002, hastie2009elements, rasmussen2006gaussian} due to its ease of implementation and interpretability. 
Moreover, it has worked successfully in numerous applications \cite{Ramakrishnan2017}. 
We note, however, that the first QML models of frontier orbital eigenvalues were presented using neural networks~\cite{Montavon2013}, and that we believe that the choice of the specific regressor is rather secondary, i.e.~our procedure could be used in combination with any other regressors just as well. 
In the following, we briefly outline the KRR methodology only for the sake of completeness.

Within KRR, the prediction of a given property $\hat{y}_i$ is given by a sum over kernel matrix elements $k_{ij} = k(\mathbf{X}_i, \mathbf{X}_j)$ multiplied by regression coefficients $\alpha_j$: 
\begin{equation}
    \hat{y}_i = \sum_{j=1}^{N_{\mathrm{tr}}} k(\mathbf{X}_i, \mathbf{X}_j) \alpha_j.
    \label{eq:y_hat}
\end{equation}
The $\mathbf{\alpha}$-coefficients are obtained by solving the following system of linear equations:
\begin{equation}
    \mathbf{y}^{\mathrm{tr}} = (\mathbf{K} + \lambda \mathbf{I} ) \boldsymbol{\alpha}.
    \label{eq:alpha}
\end{equation}
The parameter $\lambda$ is a regularization coefficient, a.k.a.~noise-level, that smooths out the noise.
However, since we are dealing with computed target values that are noiseless to machine precision, we can fix $\lambda$ to correspond to a small value such as $10^{-12}$. 
$\mathbf{I}$ is the identity matrix. 
The kernel matrix $\mathbf{K}$, for which we employ the Laplacian kernel function ($k_{ij} = \exp(-|\mathbf{X}_i - \mathbf{X}_j|_1 / \sigma)$), quantifies the similarity between any two representations of the $i$-th and $j$-th molecules. 
By virtue of this kernel matrix, each test molecule is compared to all the training molecules in order to make a prediction. 
The parameter $\sigma$ modulates the sensitivity of the kernel and is optimized via grid search cross validation within each training set in this work. 
To evaluate the performance of our method, we use the target values of the test set to calculate the mean absolute error (MAE) 
between reference and predicted values.
The logarithm of the prediction error  generally decreases linearly with the logarithm of the training set size ($\log{(E)} \propto - \log{(N_{\rm tr}})$) \cite{Cortes1994, Mueller1996}, which is shown numerically in terms of so-called learning curves.
We have employed the QML package \cite{ASChristensen2017} to perform our calculations.

\subsection{\texorpdfstring{$\Delta$}{Delta}-Machine Learning}

In $\Delta$-ML \cite{Ramakrishnan2015}, correlations between different levels of theory are exploited to obtain better predictions of properties calculated at higher levels of theory for fewer training molecules. 
We consider two levels of theory, a lower baseline, at which we know the output, and a higher target line, for which we want to obtain predictions. 
A machine is trained on the differences between the two levels. 
In other words, a QML model of a correction to the baseline model is being generated.

After that, these predictions are added to the baseline to generate estimates of the property at the higher level of theory:
\begin{equation}
    \hat{y}_i^{\mathrm{target}} = y_i^{\mathrm{baseline}} +  \sum_{j=1}^{N_{\mathrm{tr}}} k(\mathbf{X}_i, \mathbf{X}_j) \alpha_j
\end{equation}
The better the correlation between the levels of theory, the easier it is to learn the difference between them. 
In a more generalized version of this method called Multilevel-ML \cite{zaspel2018boosting}, one can exploit the correlations between more than 2 levels of theory and basis sets to improve predictions. 
In this work, we combine the SML method with $\Delta$-ML using data from the QM7b dataset, namely the ZINDO energies as baseline, and the GW energies as target.

\subsection{Selected Machine Learning}

\begin{figure}
    \centering
    \includegraphics[width=\linewidth]{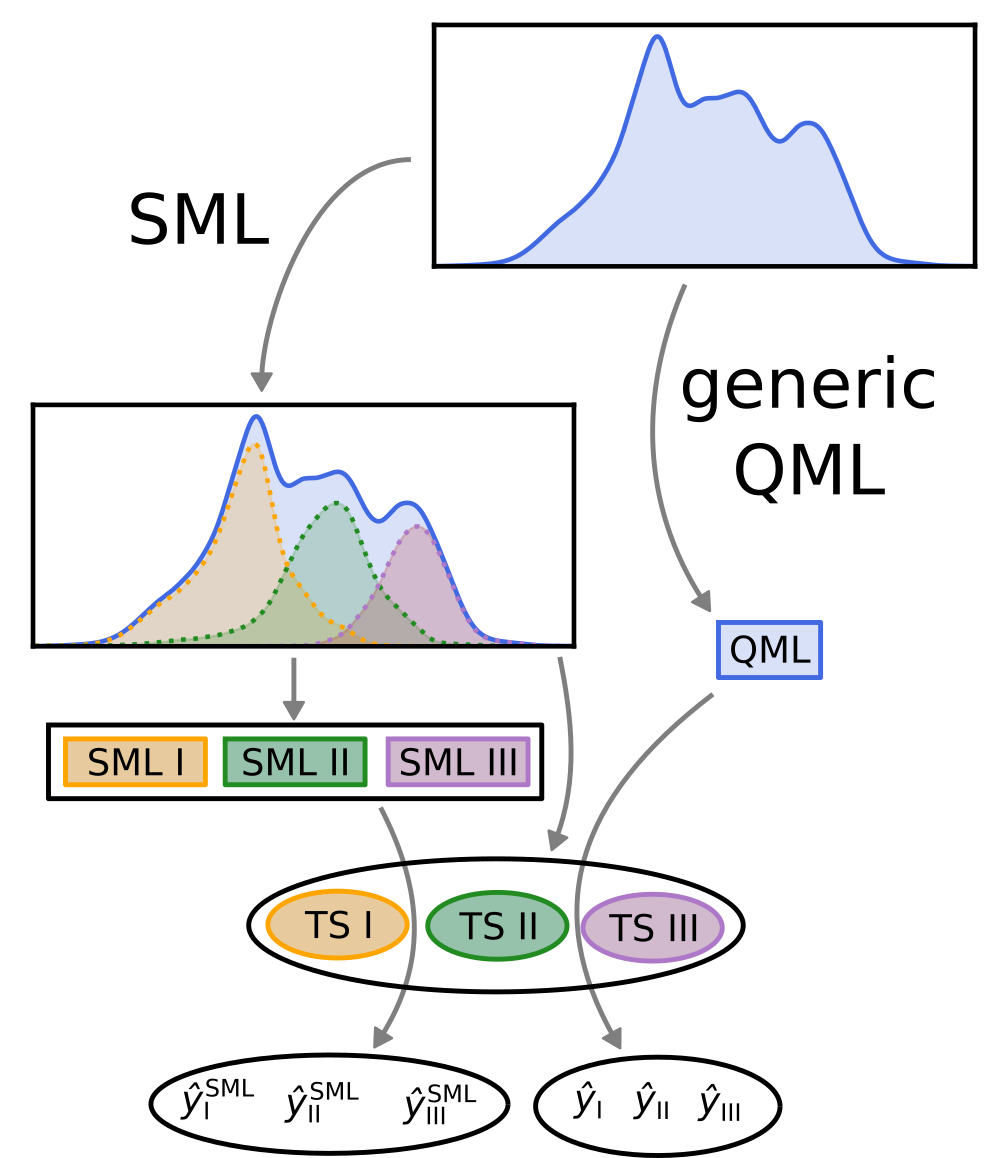}
    \caption{Visual representation of the SML method. 
    The data-set is first classified into separate classes - in our case 3. 
    For each class, a machine is trained on its molecules (models SML I, II and III) and a disjoint test set is put aside (TS I, II and III).
    In addition, one machine is trained on molecules form all across the data-set (here dubbed QML).
    Eventually, for each test set, two predictions are computed, one using generic QML ($\hat{y}_{\mathrm{I}}$, $\hat{y}_{\mathrm{II}}$ and $\hat{y}_{\mathrm{III}}$) and the other using SML ($\hat{y}_{\mathrm{I}}^{\mathrm{SML}}$, $\hat{y}_{\mathrm{II}}^{\mathrm{SML}}$ and $\hat{y}_{\mathrm{III}}^{\mathrm{SML}}$).
    These two predictions are then compared for each test set.}
    \label{fig:ml_workflow}
\end{figure}

In order to compare SML to generic QML training set selection, we proceed as visualized in Fig.~\ref{fig:ml_workflow}.
We train a model on all molecules drawn at random across the data-set (generic QML), then 3 different machines, each only with molecules from a single class (SML). 
Moreover, we generate 3 separate test sets, one for each class, while making sure that there is no overlap between any of the training and test sets. 
For each test set, we produce two predictions: one obtained from generic QML - with training molecules from all over the data-set - 
and a second one from SML - with training molecules from the corresponding class only.
We expect the prediction errors of SML to be lower than those of generic QML for each class.
By applying two different machines on exactly the same test set, their performances can be properly compared to one another.

\section{Results and Discussion} \label{sec:results}
 
\subsection{Frequency Analysis and Classification}


The graph at the bottom of panel a) in Fig.~\ref{fig:workflow} shows the frequency analysis of HOMO-LUMO gaps of QM9 molecules. 
The first row illustrates how different degrees of saturation affect the gap. 
The more unsaturated a molecule, the lower its HOMO-LUMO gap, with aromatic and fully saturated molecules having the smallest and largest gaps, respectively. 
This observation was to be expected since in unsaturated molecules, the frontier orbitals are often $\pi$-orbitals, which are closer in energy. 
The second row compares molecules with differing elemental composition. 
These distributions indicate that the composition alone is a relatively poor predictor of the location of the gap, and has thus been ignored for the classification. 
The third row illustrates the impact of the presence of  common functional group signatures including carbonyl, ester, amide bonds, and amino acids. 
Carbonyl containing compounds mostly have lower gaps, but their distribution is bimodal, with both peaks coinciding with the left and mid peak of the reference distribution. 
Therefore more specific distinctions between different types of carbonyl compounds are required.
The next two distributions suggest that amides and carboxylates (with an N- and O-atom linked to the C-atom of the carbonyl group) can be considered separately from the other carbonyl molecules, since their distributions are slightly more localized.
Albeit rather rare in the set considered, amino acids appear to be located closer to the middle peak as well. 
In conclusion, the most relevant features for the classification are saturation vs.~aromaticity, and the presence vs.~absence of a carbonyl group, since they lead to well localized sub-distributions. 
Note that HOMO-LUMO-gap distributions from the QM7b data-set at ZINDO level of theory exhibit similar structures across different groups of molecules (Fig. 3 of SI).

The graph at the bottom of panel b) of Fig.~\ref{fig:workflow} showcases the resulting classification rules of QM9 molecules used for this study.  
First, we separate all saturated molecules from the rest and assign them to one class that we call 'saturated', which corresponds to the right peak.
The remaining molecules are then subdivided into carbonyl and non-carbonyl molecules, which we further separate into aromatic and non-aromatic ones, amino acids and more specific carbonyl compounds (amides and carboxylates). 
Finally, we end up with aromatic and carbonyl molecules with more than one unsaturated bond in one class that we name 'aromatic/carbonyl', that overlaps with the left peak.
We put the remaining molecules together with amino acids and other carbonyl compounds into the last class, which we call 'single unsaturated', because most molecules of that class have only one unsaturated bond.
These rules lead to a classification which results in three molecular classes exhibiting well-behaved unimodal distributions for the data-sets considered here. 
To facilitate comparison in the following, we have numbered the classes (from left to right): class I - saturated, class II - single unsaturated, and class III - aromatic/carbonyl.
We note again that classification rules for the ZINDO gaps of QM7b (Fig.~4 of SI) are similar.
We have also performed two consistency checks of the classification with a Linear Discriminant Analysis (LDA) projection (Figs.~5-8 of SI) and a Decision Tree Classification (Fig.~10 of SI). As shown in the SI, these checks confirm the validity of our classification scheme.

In essence, our analysis suggests that the implementation of simple chemical rules enables the splitting of input molecules (solely based on structure) for a molecular data-set (here shown for QM7b and for QM9) into classes such that unimodal quantum property distributions can be obtained.

\subsection{Learning Curves}

\begin{figure}
    \centering
    \includegraphics[width=\linewidth]{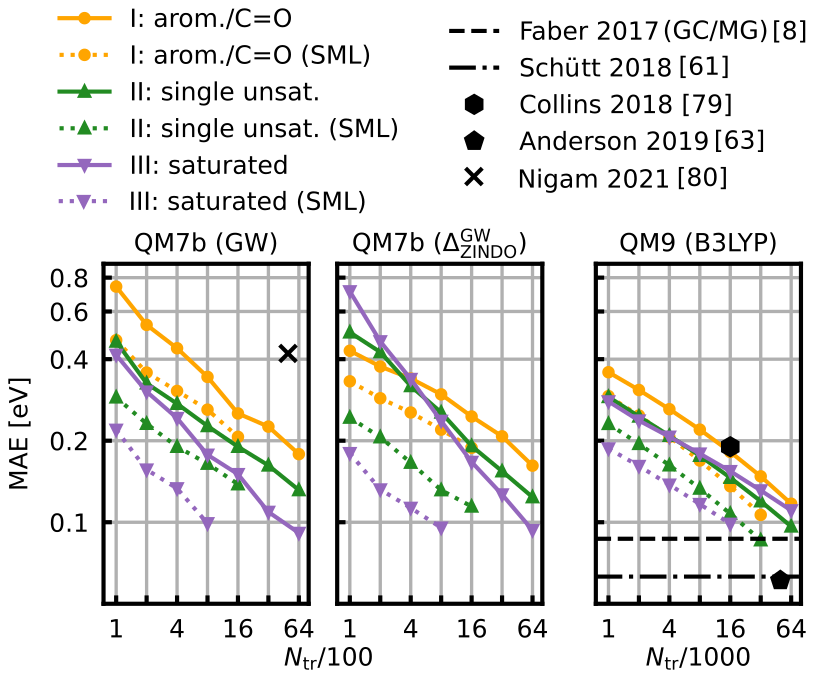}
    \caption{Learning curves for the HOMO-LUMO gap of QM7b (GW) (left), $\Delta$-ML on QM7b with ZINDO as baseline and (GW) as targetline (middle) and QM9 (B3LYP) (right). 
    The points show the MAE averaged over 10 iterations with a different selection of training set molecules each. 
    The solid lines are the learning curves obtained from training with molecules from all over the data-sets, whereas the dotted lines are obtained from Selected ML.
    Reference results from the literature \protect{\cite{Faber2017, Schuett2018, Collins2018, Anderson2019, Nigam2021}} are shown in black. 
    The results for QM9 indicated with horizontal lines were obtained with training set sizes of $\sim$110k molecules. 
    }
    \label{fig:lc_gap}
\end{figure}

The learning curves for the HOMO-LUMO gaps are presented in Fig.~\ref{fig:lc_gap}. 
They show the prediction errors w.r.t.~increasing training set size on a log-log scale.
In all cases, KRR with prior classificiation via SML (dotted lines) performs better than without classification (solid lines).
While the slopes of the learning curves remain the same, there is a significant drop for the offsets.
The largest drop can be observed for the class of saturated molecules, which is around 0.077 eV for 800 training molecules in QM7b (GW), 0.142 eV for $\Delta$-ML in QM7b and 0.055 eV in QM9.
Moreover, the prediction error for the class of saturated molecules is the lowest of all 3 classes, followed by the errors of the class of singly unsaturated molecules, and the highest errors are for the class of aromatic and carbonyl molecules.
This trend is consistent for both data-sets.
Only the learning curves of the saturated class reaches an error lower than 0.1 eV.
For QM9, SML reaches this error with 16k training samples already, while more than 64k training samples would be required without classification.
Note that this prediction error is also on par with neural network prediction errors by Faber et al.~\cite{Faber2017} which had required training on 110k training samples.
It is still far from the errors obtained by more recently developed NNs \cite{Schuett2018, Anderson2019, Liu2020}, such as, for example, a NN by Liu et al.~\cite{Liu2021} reaches errors as low as 0.032 eV, however with training sets of $\sim$100k molecules.
Linear extrapolation of SML learning curves predicts an MAE of 0.065 eV for the class of saturated molecules with 100k training molecules, close to the NN by Sch\"utt et al.~(0.63 eV) \cite{Schuett2018}. 
All in all, a prior classification systematically improves prediction errors of the HOMO-LUMO gaps.

\begin{figure}
    \centering
    \includegraphics[width=\linewidth]{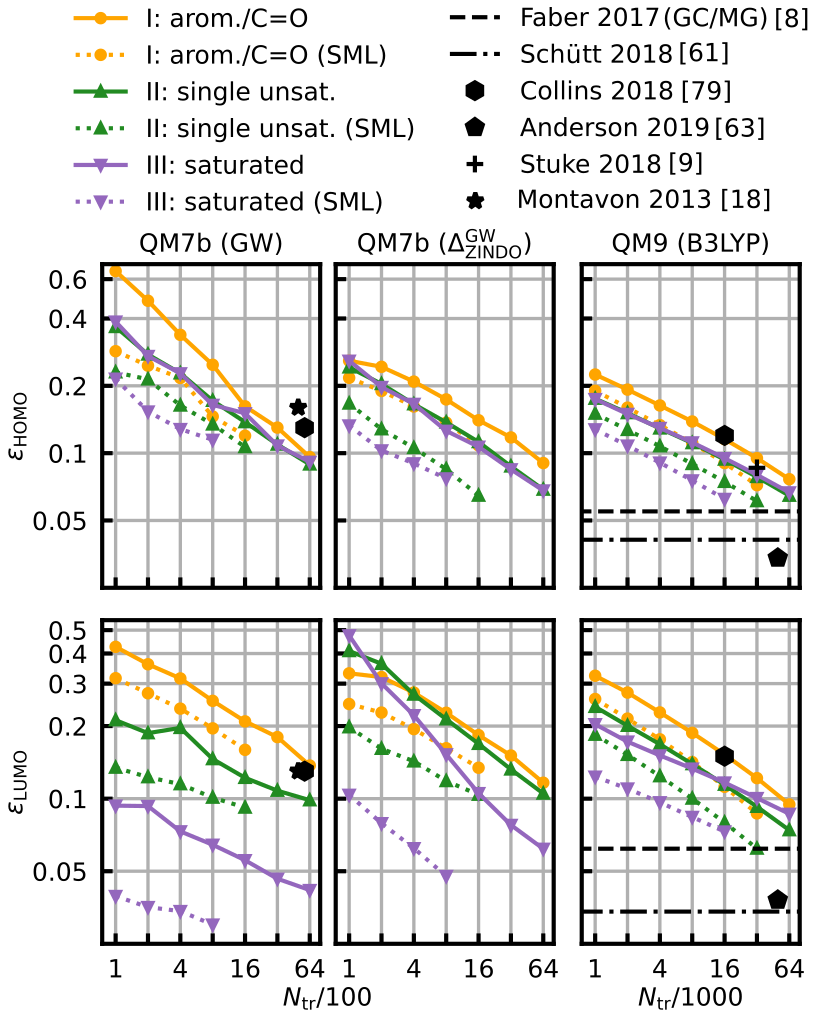}
    \caption{Learning curves for the HOMO (first row) and LUMO energies (second row).
    As in Fig.~\ref{fig:lc_gap}, results are shown for QM7b (GW), $\Delta$-ML (ZINDO to GW) and QM9 (B3LYP) from left to right and references are shown in black \protect{\cite{Faber2017, Schuett2018, Collins2018, Anderson2019, Stuke2019, Montavon2013}}.
    }
    \label{fig:lc_homo_lumo}
\end{figure}

The same model applied to the HOMO and LUMO energies results in the learning curves shown in Fig.~\ref{fig:lc_homo_lumo}. 
We can see that the learning curves generally follow the same trends as those for the HOMO-LUMO gaps: same slopes in both models, lower offsets for SML, lowest errors for the class of saturated molecules and highest errors for the class of aromatic and carbonyl molecules.
A noteworthy difference between the results for HOMO and LUMO is the extent of the improvement in the prediction errors: the errors for the HOMO energies drop less than those for the LUMO energies.
The learning curves for the LUMO energies of QM7b (GW) stand out since the error for the saturated molecules (0.030 eV) is much lower than for the other classes.
These results demonstrate that our classification can also be transferred to other related properties, such as individual HOMO and LUMO energies, even though it has been derived from gaps only.

\subsection{Scatter Plots}

\begin{figure*}
    
    \includegraphics[width=\linewidth]{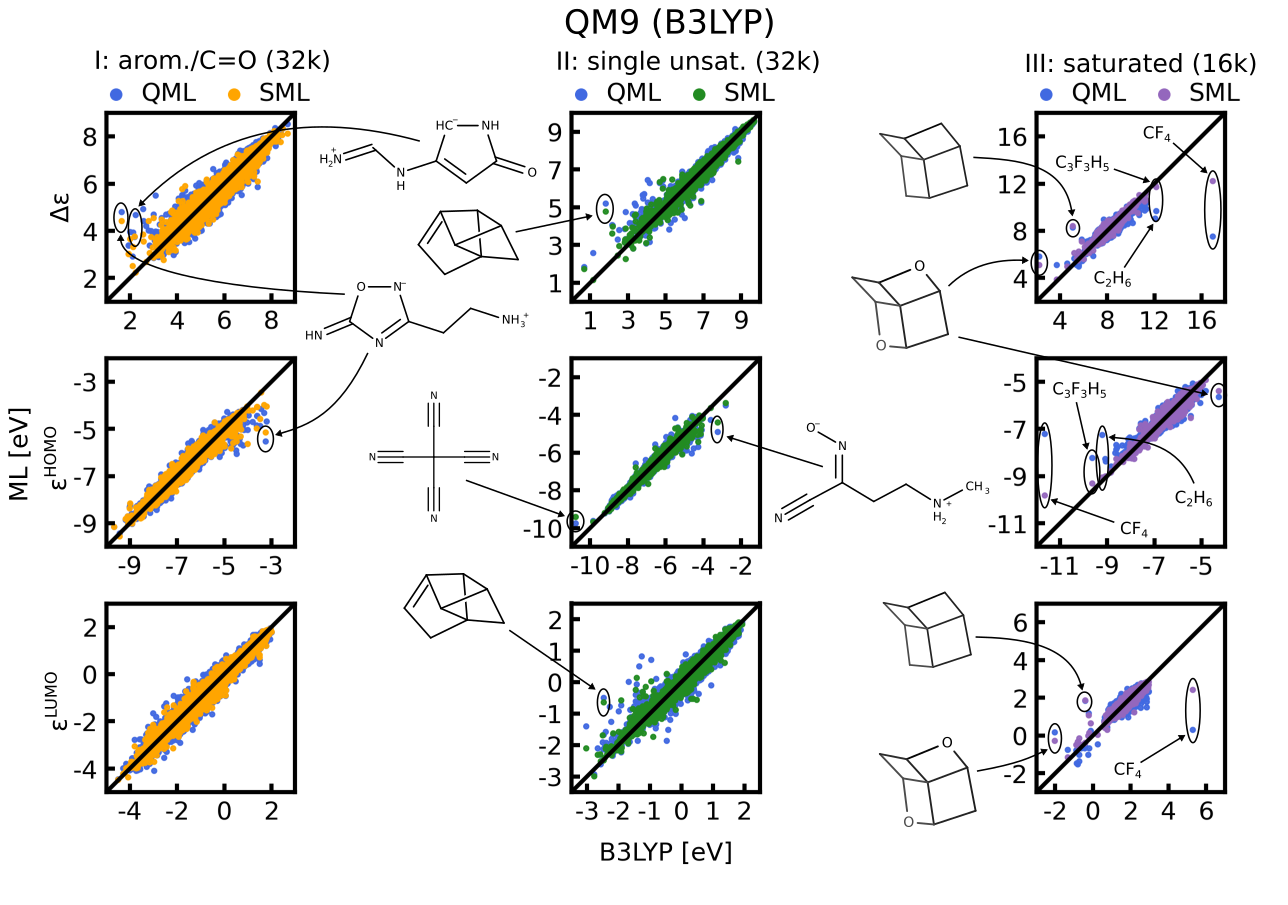}
    \caption{Scatter plots of predicted ML vs.~reference QM energies of QM9 for the largest training set size possible for each class, indicated in brackets.
    Some striking outliers are indicated in the figure, and if possible, predictions from both models (with and without classification) are highlighted for comparison.
    }
    \label{fig:scatter_qm9}
\end{figure*}

In Fig.~\ref{fig:scatter_qm9} scatter plots of prediction vs.~reference values are shown for QM9 and in Figs.~11 and 12 of the SI those for QM7b. 
The scatter plots reveal that the energies in single unsaturated and aromatic/carbonyl classes (I, and II) span a much wider range of values than the saturated molecules (class III), which explains the higher complexity and offsets of the learning curves.
Some striking outliers are labelled in the Figures. 
The most noticeable ones in the QM9 data-set are small saturated ones such as C$_2$H$_6$ or CF$_4$ (Fig.~\ref{fig:scatter_qm9}, right column). 
The HOMO and LUMO energies as well as gaps of these molecules already stand out compared to the rest, which is why ML predictions for these molecules have a large MAE.
Moreover, molecules with several rings and cage-like geometries have large prediction errors as well.
An explanation for these outliers may be that similar molecules are scarce, such that they are not necessarily well represented in the training set.
This issue could be resolved by always including such molecules in the training set.
For the other outliers highlighted we could not find a pattern that explains the large errors, we nevertheless included them for the sake of completeness.
It is worth noting that the predictions with models based on SML are in most cases closer to the exact reference values than without classification.
An exception is for instance C$_9$H$_{12}$ in the third column of Fig.~\ref{fig:scatter_qm9}, where the error from SML is larger compared to generic QML. 
In Figs.~11 and 12 of the SI more outlier examples are indicated.
In general, the predictions within the classes from SML are closer to the reference values then those from QML after random training set selection.

\subsection{Interpretation}

The systematic improvements of the prediction errors across different data-sets and properties may be explained by the reduction of effective dimensionality achieved within each of the classes.
Indeed, the lowest errors are obtained for the least diverse: the class of saturated molecules. 
Many functional groups such as carbonyl groups, aromatic rings or amides imply the presence of unsaturated bonds, indicating more chemical diversity in the other two classes.
Because of the classification, the model parameters can more easily adapt since training and test sets  contain only molecules which have increased similarity with each other. 
In other words, in order to predict the gap for an unsaturated molecule, the model did not need to account for the correlations of gaps with saturated molecules. 
In this sense, no fitting coefficients have been 'wasted' on suboptimal correlations and can contribute instead to further lower the prediction error by exploiting more effective correlations within a given class. 
Conversely, without classification, the model would be trained on many molecules with various functional groups that do not even occur in the test set, thus lowering the effective number of training molecules that are relevant for the prediction of that test set.

It is also interesting to note that in the case of the QM7b results, even though the classification was based only on the distribution of ZINDO HOMO-LUMO gaps, the prediction errors also drop for predicting the GW gaps.
As such, there seems to be a certain transferability of the classification scheme in the sense that it can be used across different levels of theory. 
But the classification is also transferable between different properties, as shown in Fig.~\ref{fig:lc_homo_lumo}.
The greater improvement for LUMO energies than for HOMO energies is likely due to correlation between gaps and LUMO energies being stronger than between gaps and HOMO energies (see Fig.~\ref{fig:gaps_hist}).

\section{Conclusion}

We have found that simple classification protocols, prior to training, can dramatically improve the data-efficiency of QML models of  HOMO-LUMO gaps in the QM9 and QM7b data sets. 
The classification is based on chemical bonding rules that allow us to define molecular classes based on structural input features alone.
Our frequency analysis reveals that the presence of  functional groups, such as aromatic rings and carbonyl groups, dominate sub-distributions of  HOMO-LUMO gaps, and can therefore be exploited for classification.
After classification, conventional kernel ridge regression based QML models afford learning curves with systematically lower offsets than without classifications. 
As a result, significantly fewer training molecules are required to reach competitive prediction errors ($\sim$0.1 eV), e.g.~16k for saturated molecules as compared to more than 64k training molecules necessary when drawing at random from QM9. 
We have also shown that our SML approach can  be applied to related individual properties, i.e.~the HOMO and LUMO energies alone.
Further analysis has indicated, that the scheme is robust across different levels of theory for the labels, i.e.~classification based on the distribution of ZINDO gaps was shown to be transferable to train more efficient QML models of GW gaps.
Comparison to $\Delta$-ML results on the same data set (QM7b) indicates that for HOMO-LUMO gaps, the classification approach presented here within offers substantially more improvement.

The additional step of prior classification alone can already lower the prediction errors in QML.
The exploitation of simple relations between molecular structure and the HOMO-LUMO gap were enough to improve learning curves consistently.
Our method can thus be considered to be a biased training set selection \cite{Browning2017,Huang2020}, since the classification leads to training sets that are specifically tailored towards the test sets.
Our results confirm that adequate training data is indeed crucial in order to obtain the best possible performance in QML.
Nevertheless, there is no universally applicable classification that would work for any data-set.
In our case, the classifications for QM9 and QM7b are only similar because they consist both of simple organic molecules, but in general, such a classification depends on the chemical space a given data-set covers, the property of interest and also the level theory at which the property is calculated.

Similar to the B3LYP and ZINDO HOMO-LUMO gaps in QM9 and QM7b respectively, other properties with multimodal distributions could also be investigated.
These could include properties related to the gap, such as excitation energies \cite{Ramakrishnan2015a}, but also properties of entirely different origin such a  highest vibrational frequencies \cite{ramakrishnan2015many} or NMR shifts. 
Similar to the gaps, one should then identify the structural features that govern these properties (well established for IR and NMR spectroscopy) in order to define molecular classes within which these distributions become unimodal and for which equal improvements in the data-efficiency of resulting QML models should be expected. 
Other future work could also involve the use of unsupervised ML methods to find new and potentially better classification rules, based on more sophisticated combinations of functional groups, or other molecular features.
In summary, simple classification rules can lead to substantial improvements in data-efficiency of QML models. 

\section*{Acknowledgements}
We acknowledge support from the European Research Council (ERC-CoG grant QML and H2020 projects BIG-MAP). 
This project has received funding from the European Union's Horizon 2020 research and
innovation programme under Grant Agreements  \#957189.
This project has received funding from the European Research Council (ERC) under the European Union’s Horizon 2020 research and innovation programme (grant agreement No. 772834).
This result only reflects the
author's view and the EU is not responsible for any use that may be made of the
information it contains.
This work was partly supported by the NCCR MARVEL, funded by the Swiss National Science Foundation. 
Some of the computational results presented have been achieved using the Vienna Scientific Cluster (VSC).
\label{sec:acknowledgements}
    
\section*{Code Availability} \label{sec:code}

The code used in this work is freely available from \url{https://github.com/b3rn4rdm/SelectedML}.

\section*{References}

\bibliography{literature}


\end{document}


\title{Supplementary Material}





\maketitle

\section{Histograms}

\begin{figure}
    \centering
    \includegraphics[width=.5\textwidth]{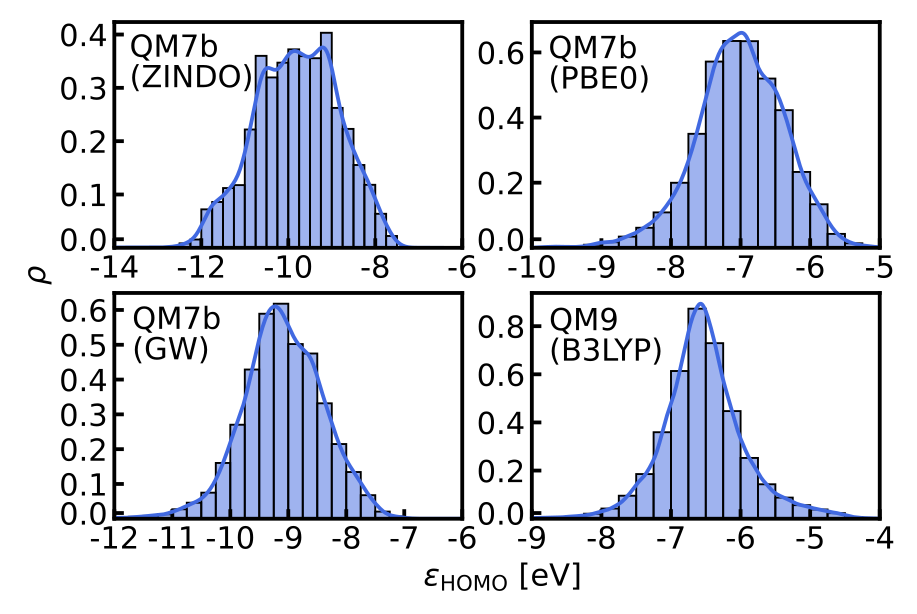}
    \caption{Histograms and KDEs of the HOMO energies of all molecules from QM7b at ZINDO, PBE0 and GW levels of theory and of all molecules from QM9 at B3LYP level of theory.}
    \label{fig:homo_hist}
\end{figure}

\begin{figure}
    \centering
    \includegraphics[width=.5\textwidth]{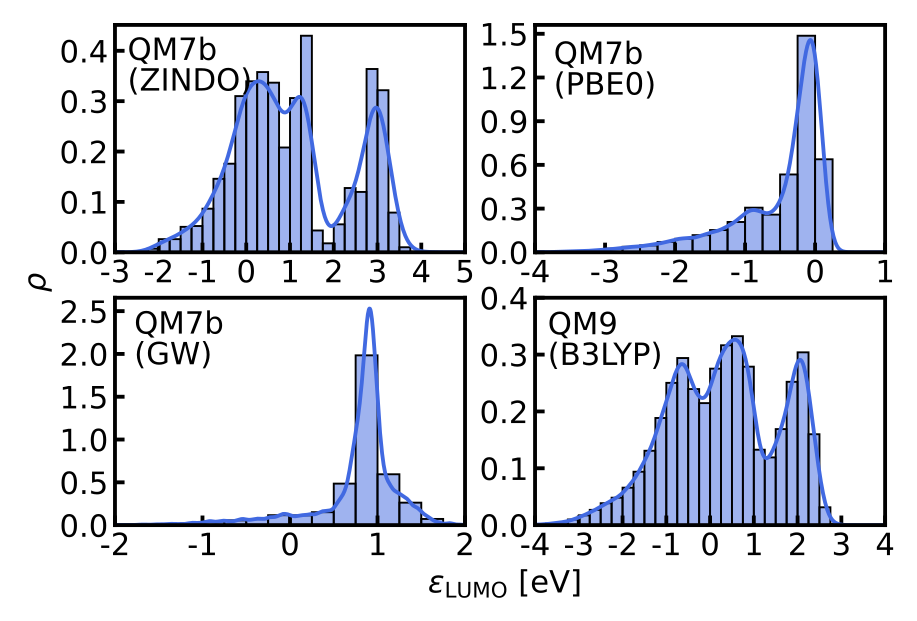}
    \caption{Histograms and KDEs of the LUMO energies of all molecules from QM7b at ZINDO, PBE0 and GW levels of theory and of all molecules from QM9 at B3LYP level of theory.}
    \label{fig:lumo_hist}
\end{figure}

In Fig.~\ref{fig:homo_hist}, histograms and KDEs are shown for the HOMO energies at the ZINDO, PBE0 and GW levels of QM7b molecules and for those at the B3LYP levels of QM9. 
The HOMO energies all follow unimodal distributions. 
Fig.~\ref{fig:lumo_hist} shows the same distributions for the LUMO energies. 
The superposition of three 3 hidden distributions similar to the distributions of gaps becomes apparent in the distributions of the ZINDO and B3LYP LUMO energies. 

\section{Frequency analysis and Classification of QM7b}

\begin{figure}
    \centering
    \includegraphics[width=.5\textwidth]{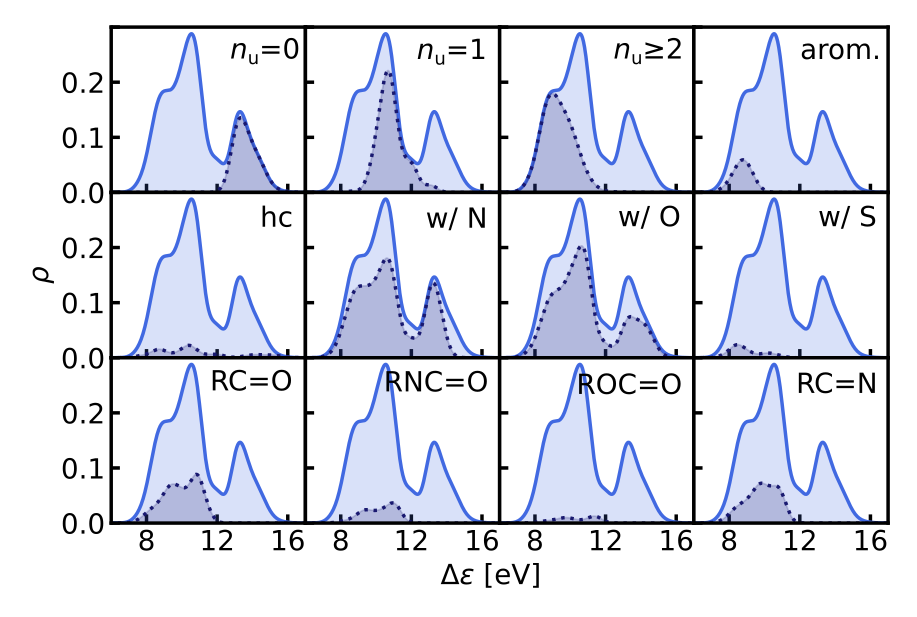}
    \caption{Frequency analysis of QM7b molecules.}
    \label{fig:qm7b_frequency_analysis}
\end{figure}

Fig.~\ref{fig:qm7b_frequency_analysis} shows the results of the frequency analysis of the ZINDO HOMO-LUMO gaps of QM7b.
As can be seen in the first row, the number of unsaturated bonds alone is already a good predictor of the gap.
Similar to QM9, the elemental composition does not have a significant influence on the gap, except maybe for sulfur.
Sulfurous compounds appear to concentrate on the lower end of the spectrum.
Carbonyl compounds are also similarly distributed as in QM9.

The final classification rules for QM7b are simpler than those for QM9. 
The separations mainly distinguish between different degrees of unsaturation.
Molecules with only saturated bonds, molecules with one unsaturated bond and molecules with more than one unsaturated bond are basically distributed between 3 disjoint classes.
The exception are nitrile compounds with no other unsaturated bond besides the carbon-nitrogen triple bond, which are assigned to the class of saturated molecules.
The resulting classification yields 3 unimodal distributions.

\begin{figure}
    \centering
    \includegraphics[width=.5\textwidth]{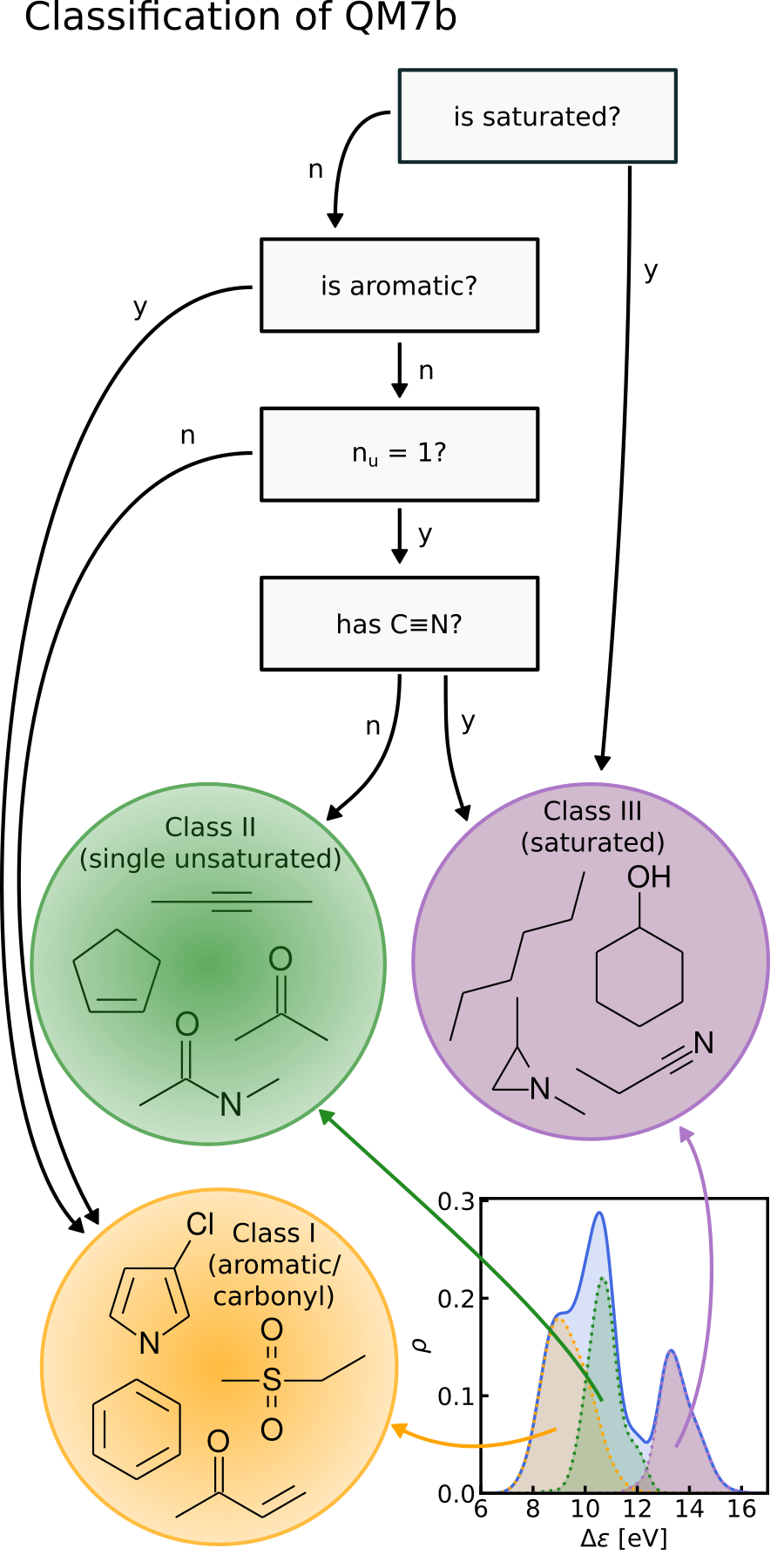}
    \caption{Workflow of the classification of QM7b molecules.}
    \label{fig:qm7b_classification}
\end{figure}

\section{LDA projection}

\begin{figure}
    \centering
    \includegraphics[width=\linewidth]{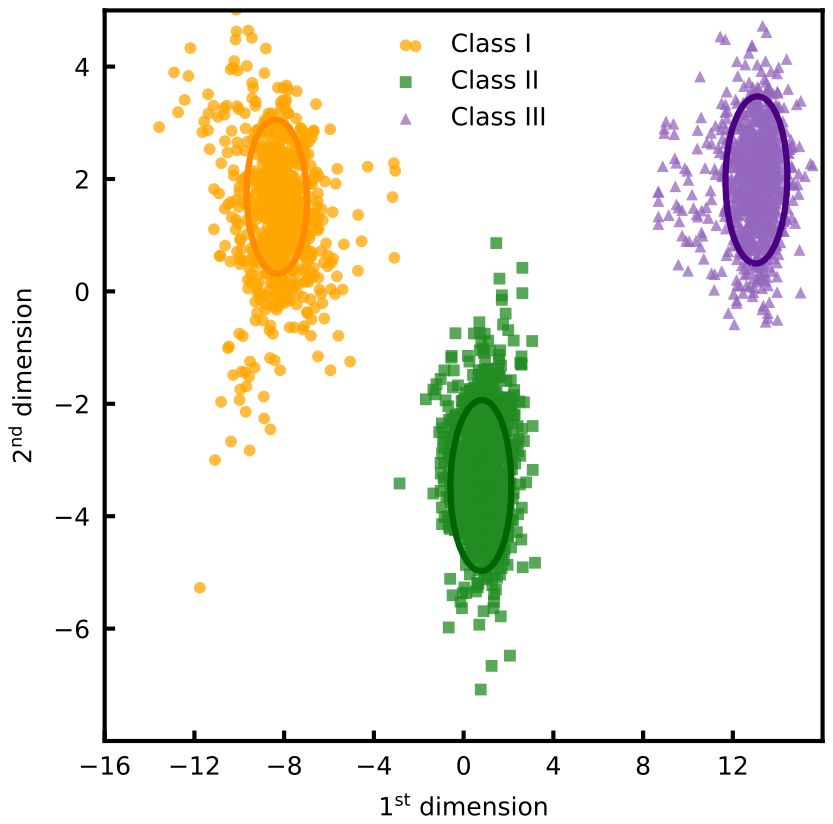}
    \caption{LDA projection of QM7b with SLATM with contour lines showing the general shape of the class distributions.}
    \label{fig:qm7b_lda_slatm}
\end{figure}

\begin{figure}
    \centering
    \includegraphics[width=\linewidth]{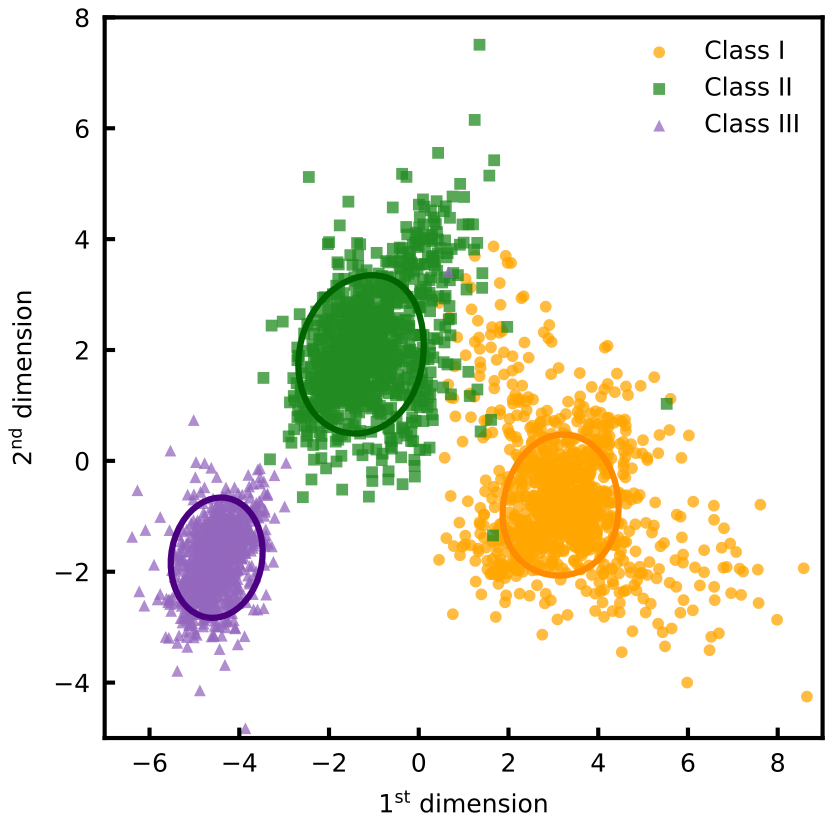}
    \caption{LDA projection of QM9 with SLATM with contour lines showing the general shape of the class distributions. .}
    \label{fig:qm9_lda_slatm}
\end{figure}

As a consistency check of our classification rules, we perform a dimensionality reduction of our data with a Linear Discriminant Analysis (LDA) projection \cite{hastie2009elements}.
LDA is a supervised ML model that, given some inputs and corresponding class labels, seeks to find the best possible linear separation of the inputs between the classes.
Furthermore, it can be used to project the input data onto a subspace that maximizes the variance between classes while minimizing the variance within classes.
This projection thus provides an intuitive visualization of how well the inputs are separable by a linear model, given our class labels.
As can be seen in Figs.~\ref{fig:qm7b_lda_slatm} and \ref{fig:qm9_lda_slatm}, our 3 classes form easily separable classes.
In the case of QM7b, the classes are even perfectly separable.

\begin{figure}
    \centering
    \includegraphics[width=\linewidth]{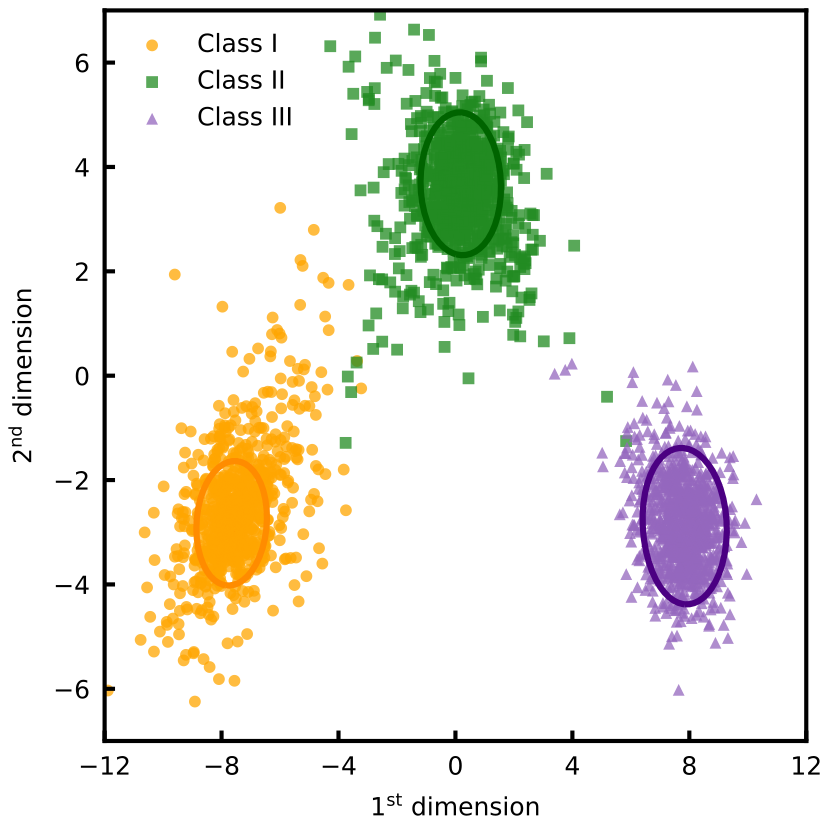}
    \caption{LDA projection of QM7b with SLATM using labels from Gaussian Mixture Model with contour lines showing the general shape of the class distributions.}
    \label{fig:qm7b_lda_slatm_gm}
\end{figure}

\begin{figure}
    \centering
    \includegraphics[width=\linewidth]{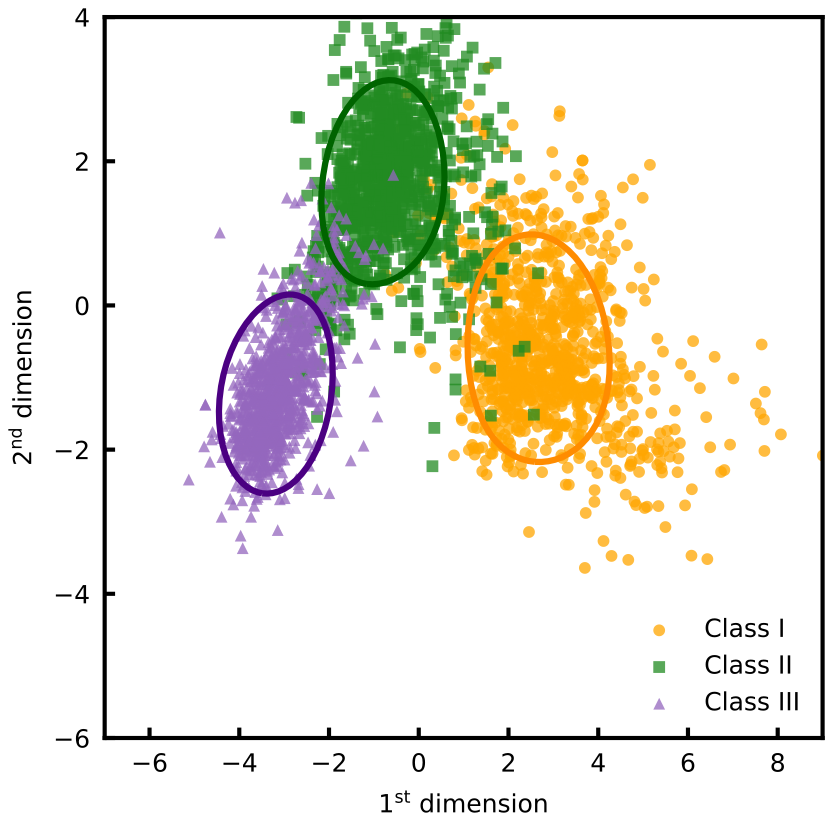}
    \caption{LDA projection of QM9 with SLATM using labels from Gaussian Mixture Model with contour lines showing the general shape of the class distributions. The same 1000 randomly selected molecules as in Fig.~\ref{fig:qm9_lda_slatm} were used.}
    \label{fig:qm9_lda_slatm_gm}
\end{figure}

For comparison, we also used a Gaussian Mixture Model (GMM), an unsupervised ML model, to generate a set of class labels from the HOMO-LUMO gap distribution.
As done above, we used LDA to project the inputs onto a subspace the best separates the classes given the new labels.
Figs.~\ref{fig:qm7b_lda_slatm_gm} and \ref{fig:qm9_lda_slatm_gm} show the resulting projections.
One can clearly see that there is a more overlap between the clusters than for our class labels.
An explanation is that the GMM only takes into account the values of the HOMO-LUMO gaps, but ignores the input structures. 
These results confirm that our classification rules lead to well separable classes.
For both the LDA projection and GMM classification, the implementations from \texttt{scikit-learn} \cite{Pedregosa2011} were used.
  
\section{Decision Trees}

\begin{figure}
    \centering
    \includegraphics[width=.5\textwidth]{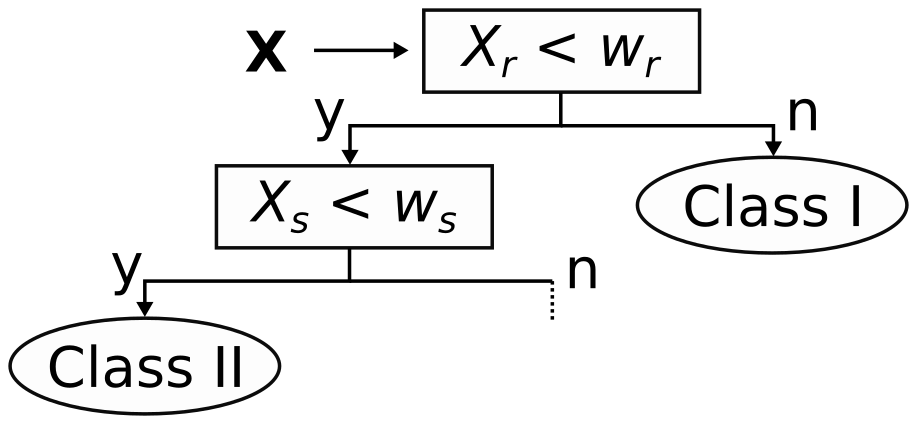}
    \caption{Schematic representation of the classification done by a DT classifier. Each node represents a decision that checks one value of the representation vector. Eventually, a leaf node is reached with the class estimate for the input.}
    \label{fig:decision_tree_scheme}
\end{figure}

We also let a supervised ML classifier learn the labels of our classification. 
In this classification, the target values are the class labels the molecules belong to.
We found a Decision Tree Classifier (DTC) as implemented in \texttt{scikit-learn} \cite{Pedregosa2011} to yield good results for our problem. 
In a DTC, a sequence of binary decisions is applied to an input $\mathbf{X}_i$ until the algorithm can conclusively tell to which class it belongs (see Fig.~\ref{fig:decision_tree_scheme}). 
All the decisions are represented by nodes that eventually lead to a leaf node which tells the estimated class label. 
The classifier's performance is measured by its score: the proportion of correctly assigned labels to the total number of samples in the test set. 

\begin{figure}
    \centering
    \includegraphics[width=.5\textwidth]{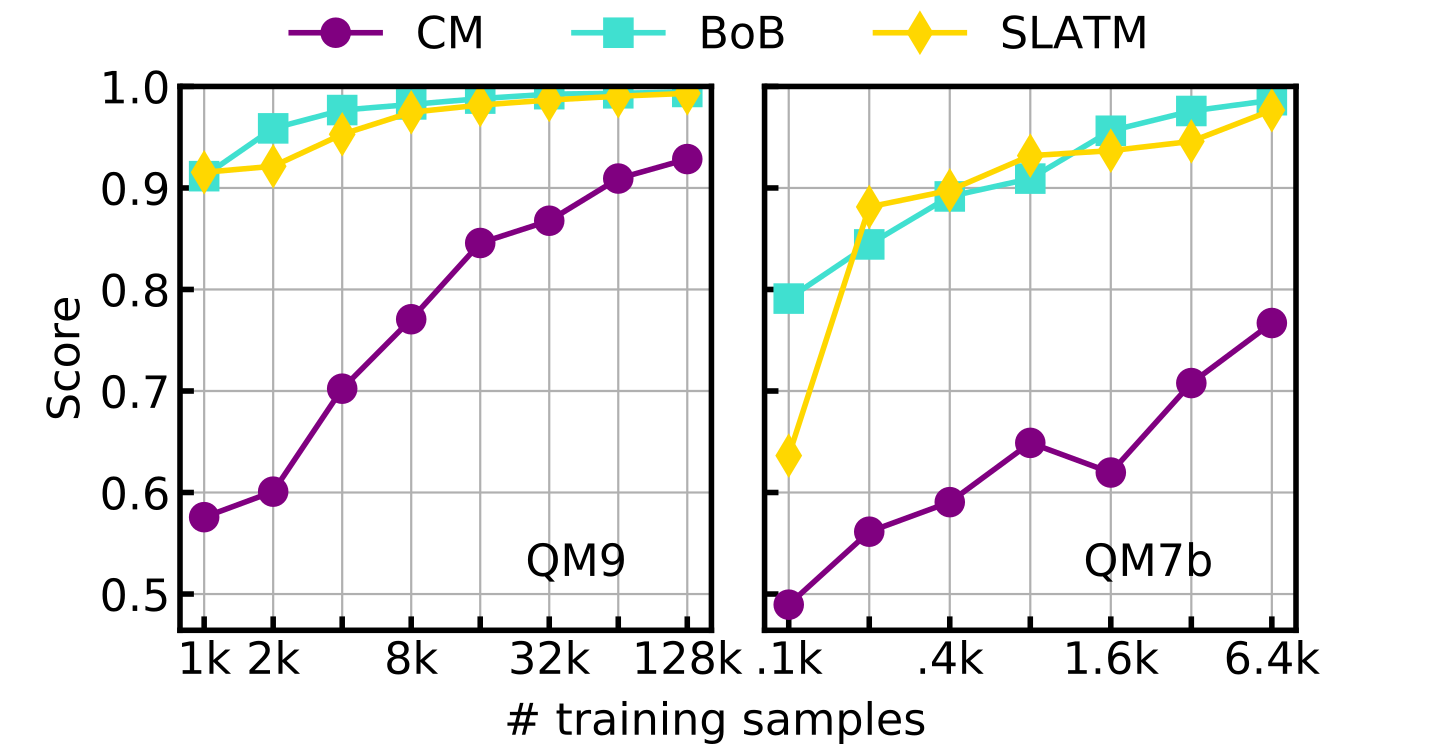}
    \caption{Learning curves with the score of Decision Tree Classification for three representations: CM, BoB and SLATM.}
    \label{fig:score}
\end{figure}

The results from the DT classification using the CM \cite{Hansen2013}, BoB \cite{Hansen2015} and SLATM \cite{Huang2020} representations for both QM9 and QM7b are shown in Fig.~\ref{fig:score}. 
The score improves with increasing training set size, but BoB and SLATM are clearly better suited for the classification the CM. 
With the former representations, the DT classifier starts with a score above 0.9 already for as few 1000 training samples and reaches a score of 0.99 with 124k training samples. 
With the CM however, the DTC requires at least 64k training samples to reach a score over 0.9. 
This result demonstrates that our classification rules can easily be learned by ML algorithms.

\section{Scatter Plots}

\begin{figure*}
    \centering
    \includegraphics[width=\textwidth]{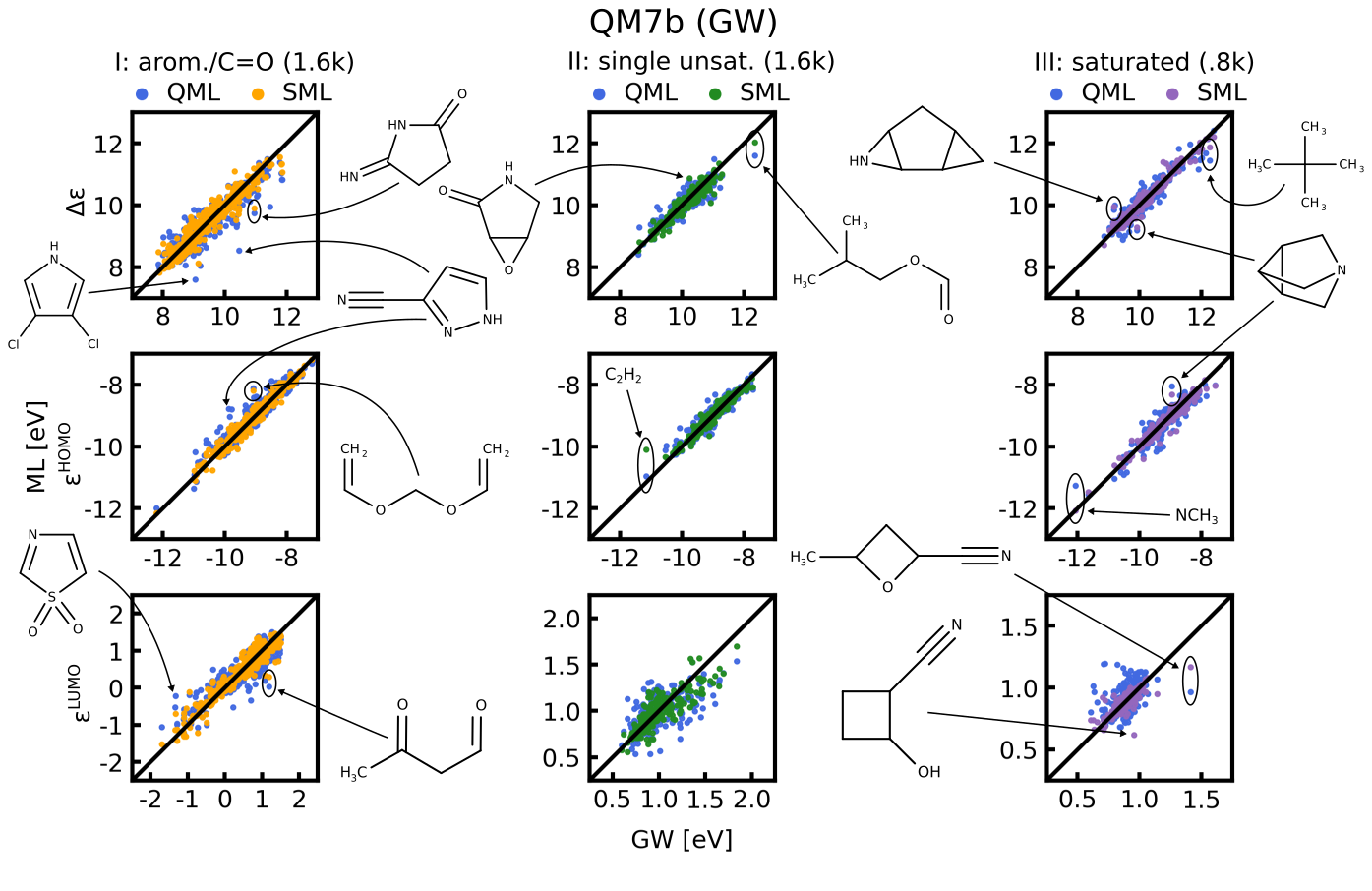}
    \caption{Scatter plots of predicted ML vs.~reference QM energies of QM7b (GW) for the largest training set size possible for each class, with outliers.}
    \label{fig:scatter_qm7b_gw}
\end{figure*}

\begin{figure*}
    \centering
    \includegraphics[width=\textwidth]{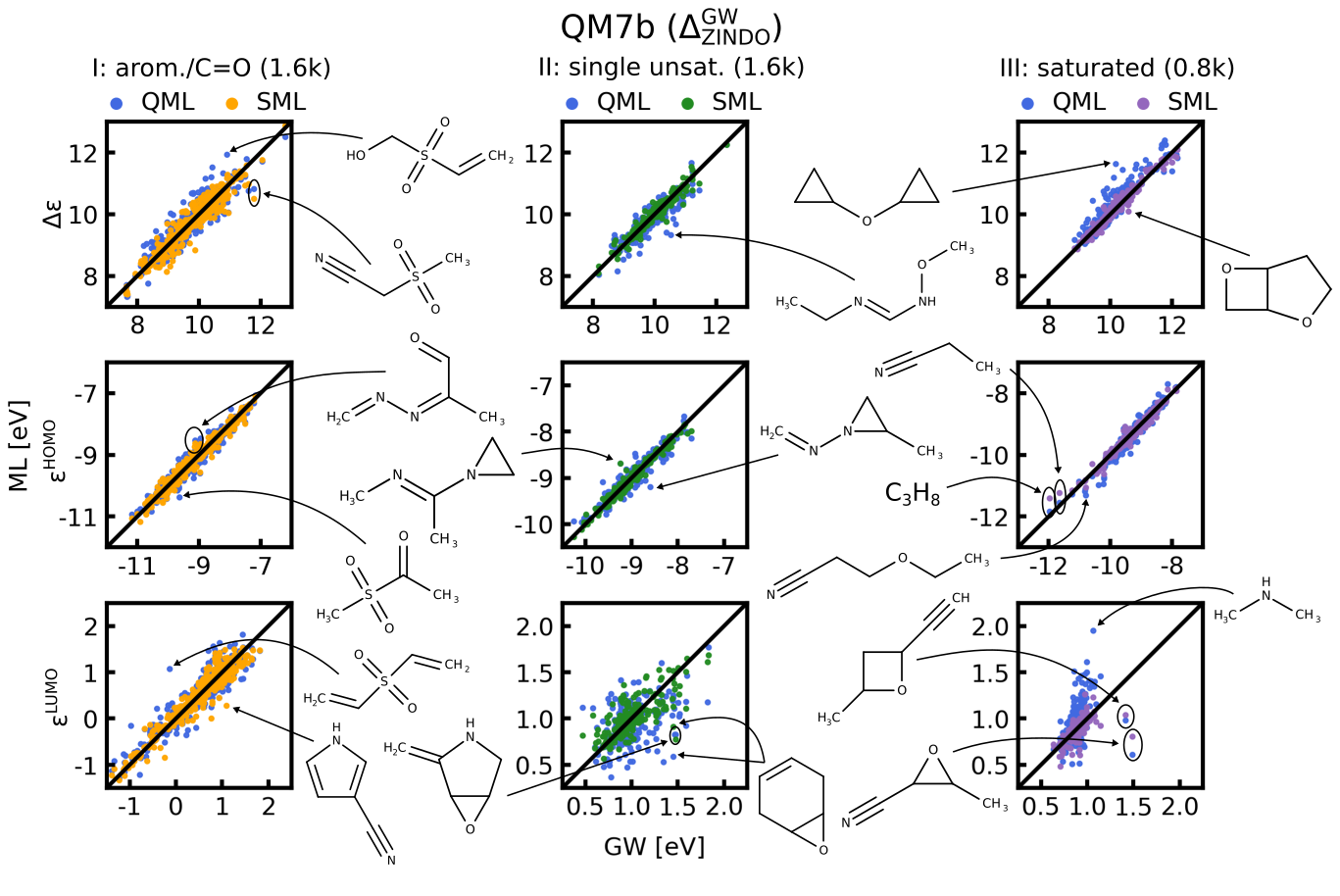}
    \caption{Scatter plots of predicted ML vs.~reference QM energies of QM7b ($\Delta_{\mathrm{ZINDO}}^{\mathrm{GW}}$) for the largest training set size possible for each class, with outliers.}
    \label{fig:scatter_qm7b_delta}
\end{figure*}

Figs.~\ref{fig:scatter_qm7b_gw} and \ref{fig:scatter_qm7b_delta} show the scatter plots of ML predictions vs.~QM reference values for QM7b (GW) and $\Delta$-ML. 
It is noticeable that there appear to be more cases in the QM7b data set where SML fails to produce better predictions than in QM9.
We assume that this result is due to the fact that the classification is based on the ZINDO HOMO-LUMO-gaps, while the predictions are for properties computed at GW level of theory.

\bibliography{si_literature}





